\documentclass[twocolumn, superscriptaddress, preprintnumbers,amsmath,amssymb, showpacs]{revtex4}

\allowdisplaybreaks
\allowdisplaybreaks[4]
\usepackage{graphicx,latexsym}
\usepackage{multirow,array,booktabs}
\usepackage{subfigure}
\usepackage[figuresright]{rotating}
\usepackage{color}
\usepackage{CJK}
\usepackage{txfonts}
\usepackage{dcolumn}
\usepackage{bm}
\usepackage[dvipdfm,
            pdfstartview=FitH,
            CJKbookmarks=true,
            bookmarksnumbered=true,
            bookmarksopen=true,
            colorlinks=true,
            pdfborder=001,
            linkcolor=blue,
            anchorcolor=blue,
            citecolor=blue
            ]{hyperref}

\begin{document}

\title{Covariant density functional theory for nuclear chirality in $^{135}$Nd}

\author{J. Peng}\email{jpeng@bnu.edu.cn}
\affiliation{Department of Physics, Beijing Normal University,
Beijing 100875, China}

\author{Q. B. Chen}\email{qbchen@pku.edu.cn}
\affiliation{Physik-Department, Technische Universit\"{a}t
M\"{u}nchen, D-85747 Garching, Germany}

\begin{abstract}

The three-dimensional tilted axis cranking covariant density functional
theory (3D-TAC CDFT) is used to study the chiral modes in $^{135}$Nd.
By modeling the motion of the nucleus in rotating mean field as
the interplay between the single-particle motions of several valence
particle(s) and hole(s) and the collective motion of a core-like part, a classical
Routhian is extracted. This classical Routhian gives qualitative agreement
with the 3D-TAC CDFT result for the critical frequency corresponding to
the transition from planar to aplanar rotation. Based on this investigation
a possible understanding of tilted rotation appearing in a microscopic
theory is provided.

\end{abstract}


\maketitle


Triaxial nucleus, shaped like a kiwi with all three principal
axes unequal, has drawn considerable attention over the years.
While being rare in the ground state~\cite{Moller2006PRL}, it
becomes more common at high spin~\cite{Werner1995ADNDT}.
For high spin, the orientation of angular momentum vector
relative to the triaxially deformed density distribution
becomes a very important concept~\cite{Frauendorf2001RMP}.
The angular momentum may align along one of principal axes
(principal axis rotation), lie in one of the principal
planes defined by two principal axes (planar rotation),
and deviate from the principal planes (aplanar rotation).
Note that the second and the third motions are also called as
tilted rotation.

For tilted rotation, its physical interpretation was given in
Ref.~\cite{Frauendorf1997NPA} by studying the model system
of one $h_{11/2}$ particle and one $h_{11/2}$ hole coupled to
a triaxial rotor with the irrotational flow type of moments of
inertia (MoIs). In detail, the high-$j$ particle/hole tends to align
along the short/long ($s$-/$l$-) axis because its torus-like/dumbbell-like
density distribution has the maximal overlap with the triaxial core
in the medium-long/medium-short ($ml$-/$ms$-) plane.
The triaxial rotor tends to align along the medium ($m$-)
axis since it is of the largest MoI. As a consequence,
the total angular momentum will firstly lie in the $sl$-plane
and with increasing spin be tilted towards the $m$-axis,
i.e., showing a transition from planar to aplanar rotation.

The occurrence of aplanar rotation manifests itself as the observation
of a pair of nearly degenerated $\Delta I=1$ bands with the same
parity in the laboratory. They are called as chiral doublet
bands~\cite{Frauendorf1997NPA}, since in the aplanar rotation the mutually
perpendicular angular momenta of the valence particle, valence hole,
and rotor can be arranged to form two systems which differ by chirality,
i.e., a left- and a right- handed system. Correspondingly, the
excitation modes based on the aforementioned planar and aplanar
rotations are called as chiral vibration and chiral rotation,
respectively.

As is well-known, the chirality is a phenomenon existing commonly in nature,
such as, the spiral arm of the nebula, the spirals of snail shells, and
the handedness of amino acids. Therefore, the propose~\cite{Frauendorf1997NPA}
and observation~\cite{Starosta2001PRL} of chirality at nuclear high spin state
caused quite a stir among nuclear structure physics~\footnote{See e.g.,
\url{https://www.sciencemag.org/news/2001/02/nuclei-crash-through-looking-glass}.},
and has intrigued lots of investigations from both theoretical and experimental
aspects. Up to now, more than 50 chiral nucleus candidates spread in the
mass regions of $A \sim 80$, $100$, $130$, and $190$ have been reported,
see, e.g., data tables~\cite{B.W.Xiong2019ADNDT}.

As mentioned above, the nuclear chirality was proposed based on the
study of angular momentum vector geometry of a phenomenological
model system with a particle and a hole coupled to a triaxial
rotor~\cite{Frauendorf1997NPA}. This model system reveals the
duality of the single-particle and collective properties of atomic
nuclei and has been used extensively in the development of
(triaxial) particle rotor model (PRM)~\cite{Bohr1976RMP, Bohr1975}.
By the newly developed many-particle-many-hole PRM (e.g.,
Refs.~\cite{J.Peng2003PRC, Koike2004PRL, S.Q.Zhang2007PRC,
B.Qi2009PLB, Lawrie2010PLB, Starosta2017PS, Q.B.Chen2018PLB,
Q.B.Chen2019PRC, Y.Y.Wang2019PLB, J.Peng2019PLB, J.Peng2020PLB}),
the understanding of the nuclear chirality is deeply entrenched.

To verify the prediction by the model system, microscopic three-dimensional
tilted axis cranking (3D-TAC) calculations in a hybrid Woods-Saxon
and Nilsson model combined with the shell correction method were
carried out~\cite{Dimitrov2000PRL}. In the 3D-TAC calculation,
with the mean field approximation for the two body interaction, all of the nucleons
are treated on the same footing. It allows also for an arbitrary orientation
of the angular momentum vector in the intrinsic frame. The solution for each
rotational frequency $\omega$ is obtained self-consistently by minimizing the
total Routhian surface, which is the energy surface of the rotating nucleus in
the intrinsic frame, with respect to the orientation of angular velocity vector
$\bm{\omega}=\omega(\sin\theta\cos\varphi, \sin\theta\sin\varphi,\cos\theta)$.
Here, usually, $\theta$ is the polar angle between the $\bm{\omega}$ and $l$-axis,
and $\varphi$ is the azimuthal angle between the projection of $\bm{\omega}$
onto the $ms$-plane and the $s$-axis. The authors of Ref.~\cite{Dimitrov2000PRL}
found that when cranking a realistic system with triaxially deformation
and high-$j$ particle-hole configuration, $\theta \neq 0$, $\pm \pi/2$, $\varphi=0$,
$\pm \pi/2$ for planar rotation at low $\omega$, and $\theta \neq 0$, $\pm \pi/2$,
$\varphi \neq 0$, $\pm \pi/2$ for aplanar rotation above a certain value of
$\omega$ (denoted as $\omega_{\textrm{crit}}$).

Subsequently, an analytical formula of $\omega_{\textrm{crit}}$, called as the
critical frequency, was derived in Ref.~\cite{Olbratowski2004PRL}. In the
derivation, using the aforementioned model system, the particle and
hole are further assumed to align rigidly along the $s$- and $l$- axis,
respectively. Further self-consistent calculations by the 3D-TAC
Skyrme-Hartree-Fock method show that $\omega_{\textrm{crit}}$
depends on the effective interaction used~\cite{Olbratowski2004PRL, Olbratowski2006PRC}.
Very recently, 3D-TAC calculations based on covariant density functional
theory (3D-TAC CDFT)~\cite{P.W.Zhao2017PLB} also support
the existence of $\omega_{\textrm{crit}}$.

In the 3D-TAC calculations, the model system that valence nucleons
coupled to a triaxial rotor is not necessary to be assumed a priori,
since all of the nucleons are treated on the same footing in the rotating
mean field. Then a straightforward question comes: which part of nucleon(s)
drives the total nuclear system to exhibit an aplanar rotation? Inspired
by the entrenched picture in model system, we might wonder whether there
exists really a hidden collective core-like part and this ``core'' plays
a role driving the rotational axis toward to the $m$-axis to stimulate the
aplanar rotation. In this work, we try to figure out this core-like part and based
on it provide a possible understanding of tilted rotation appearing in the
microscopic 3D-TAC calculations.

As an example, the chiral modes in nucleus $^{135}$Nd will be
studied. The reasons for studying $^{135}$Nd are as follows.
Experimentally, this nucleus is a possible candidate nucleus with
multiple chiral doublet (M$\chi$D) phenomenon~\cite{J.Meng2006PRC,
Ayangeakaa2013PRL} as two pairs of chiral doublet bands were reported.
One of them (labeled as bands D5 and D6 in the current work), built on
the configuration $\pi h^2_{11/2}\otimes \nu h^{-1}_{11/2}$ (labeled as
config2), is the first reported chiral doublet bands of
odd-$A$ nuclei~\cite{S.Zhu2003PRL} and with rare lifetime
measurement results~\cite{Mukhopadhyay2007PRL}. The other one
(labeled as bands D3 and D4), built on the configuration
$\pi [h^1_{11/2}(gd)^1] \otimes \nu h^{-1}_{11/2}$ (labeled as config1),
is just established very recently~\cite{B.F.Lv2019PRC}. Theoretically, $^{135}$Nd
has drawn attentions from various differen kinds of models, such as
the microscopic 3D-TAC model based on hybrid of Woods-Saxon and Nilsson
potential combined with the shell correction method~\cite{S.Zhu2003PRL}
and the 2D-TAC CDFT~\cite{P.W.Zhao2015PRC}, the beyond cranking mean field
random phase approximation (RPA)~\cite{Mukhopadhyay2007PRL}, the
algebraic interacting boson-fermion model (IBFM)~\cite{Brant2009PRC},
and the phenomenological PRM~\cite{B.Qi2009PLB, B.F.Lv2019PRC}.
Therefore, the study of $^{135}$Nd is a matter of general interest
and has double meaning of experimental and theoretical aspects.
In this work, the 3D-TAC CDFT will be further used to study the
chirality in this nucleus.

The covariant density functional theory (DFT) based on the mean
field approach has played an important role in a fully microscopic
and universal description of a large number of nuclear
phenomena~\cite{Reinhard1996RPP, Ring1996PPNP, Serot1997IJMPE,
Vretenar2005PhysRep, J.Meng2006PPNP}. In order to describe nuclear
rotation, covariant DFT has been extended with the cranking
mode~\cite{Konig1993PRL, Madokoro2000PRC, Peng2008PRC, Zhao2011PLB,
P.W.Zhao2017PLB}. Using these extended cranking covariant DFT, lots
of rotational excited states has been described well, such as
superdeformed bands~\cite{Konig1993PRL},
magnetic~\cite{Madokoro2000PRC,Peng2008PRC, Zhao2011PLB,
Steppenbeck2012PRC,Yu2012PRC} and antimagnetic
rotations~\cite{Zhao2011PRL, Zhao2012PRC, P.Zhang2014PRC,
Peng2015PRC}, etc. For chirality in nuclei, M$\chi$D phenomenon has
been suggested and investigated based on constrained triaxial
covariant DFT calculations~\cite{J.Meng2006PRC, J.Peng2008PRC,
Yao2009PRC, J.Li2011PRC, J.Li2018PRC, B.Qi2018PRC, J.Peng2018PRC}.
In particular, the newly developed three-dimensional tilted axis
cranking (3D-TAC) covariant DFT has been successfully applied for the
chirality in $^{106}$Rh~\cite{P.W.Zhao2017PLB},
$^{136}$Nd~\cite{Petrache2018PRC} and
$^{106}$Ag~\cite{P.W.Zhao2019PRC}.


Our microscopic calculations are performed using 3D-TAC
CDFT~\cite{P.W.Zhao2017PLB, P.W.Zhao2019PRC} with the effective
point-coupling interaction PC-PK1~\cite{P.W.Zhao2010PRC}, and a
three-dimensional Cartesian harmonic oscillator basis with ten major
shells are used to solve the Dirac equation. The pairing correlation
is neglected in the current investigation, but one has to bear in
mind it could have some influences on the critical
frequency~\cite{Olbratowski2004PRL} and the descriptions of total
angular momentum and $B(M1)$ values~\cite{P.W.Zhao2015PRC}.


As mentioned above, in 3D-TAC calculations, the orientation of the
angular velocity $\bm{\omega}$ with respect to the three principal
axes $(\theta, \varphi)$ is determined in a self-consistent way by
minimizing the total Routhian. In these minima with $(\theta_{\min},
\varphi_{\min})$, the deformation parameters ($\beta$, $\gamma$) are
about $(0.23, 21^{\circ})$ for config1 and $(0.24, 22^{\circ})$ for
config2, respectively. They change within $\Delta \beta=0.02$ and
$\Delta \gamma =3^\circ$, exhibiting a stability with the rotation.
The $\theta_{\min}$ varies from 50$^\circ$ to 68$^\circ$ for config1
and from $63^\circ$ to $66^\circ$ for config2 driven by rotation.

\begin{figure}[!ht]
\includegraphics[width=8.5 cm]{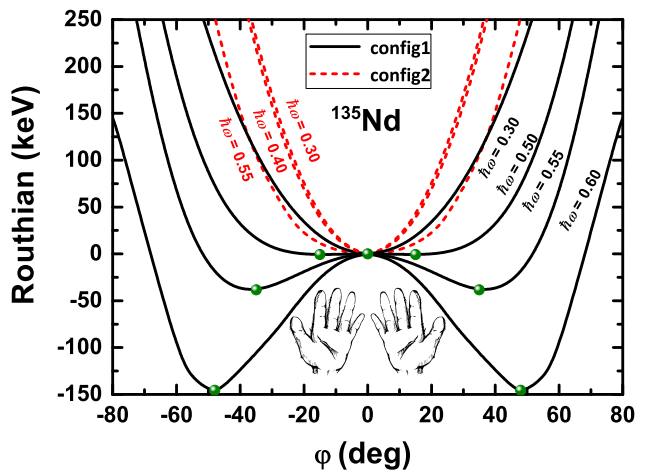}
   \caption{Total Routhian curves as functions of the azimuth angle $\varphi$
   calculated by 3D-TAC CDFT for configurations $\pi [h^1_{11/2}(gd)^1]\otimes \nu h^{-1}_{11/2}$
   (config1) and $\pi h^2_{11/2}\otimes \nu h^{-1}_{11/2}$ (config2) in $^{135}$Nd.
   All curves are normalized at $\varphi=0$. The minima of the curves are denoted
   by the balls.}
   \label{Routhians}
\end{figure}

Since the sign of $\varphi$ can be used to characterize the chirality
of the rotating system~\cite{Q.B.Chen2013PRC}, here we can only focus
on how does the total Routhian behave with $\varphi$. By minimizing
the total Routhians with respect to $\theta$ for given $\varphi$,
the total Routhian curves are shown in Fig.~\ref{Routhians} for two
configurations at several typical rotational frequencies. One can see
that all the curves are, analogy to the schematic picture of left-
and right- hand, symmetrical with the $\varphi=0$ line. This means
that the two chiral configurations with $\pm |\varphi|$ for a
given $\theta$ are identical on the energy.

For both configurations, the Routhian curves are rather steep at low
$\omega$, and become softer with respect to $\varphi$ with increasing
$\omega$. The $\varphi_{\min}=0$ corresponds to a planar rotation in the
$sl$-plane for the yrast band. At slightly larger energy, the angular
momentum $\bm{J}$ in fact could, according to the beyond TAC mean field
approximation investigations~\cite{Mukhopadhyay2007PRL, Almehed2011PRC,
Q.B.Chen2013PRC, Q.B.Chen2016PRC}, execute harmonic oscillation with
respect to the $sl$-plane and enter into the left- and right- sector
back and forth. This motion, generating the yrare band, is called as
chiral vibration~\cite{Starosta2001PRL}.

\begin{figure}[!ht]
\includegraphics[width=8.5 cm]{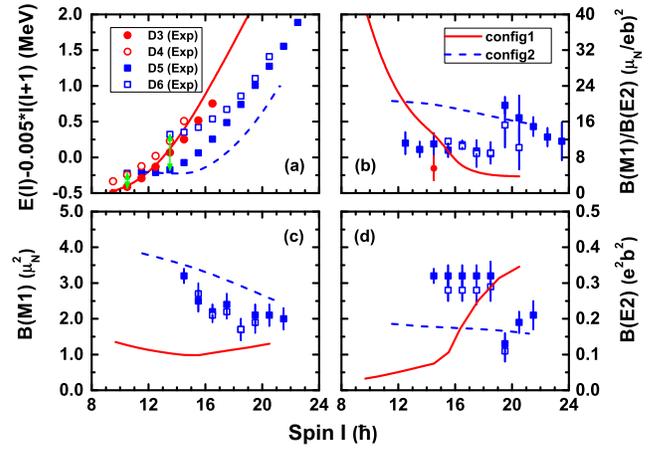}
   \caption{Calculated results by 3D-TAC CDFT in comparisons with
   the available data~\cite{S.Zhu2003PRL, B.F.Lv2019PRC, Mukhopadhyay2007PRL}
   of doublet bands D3 and D4 and bands D5 and D6 in $^{135}$Nd:
   (a) energies minus a common rotor contribution, (b) $B(M1)/B(E2)$ ratios,
   (c) $B(M1)$, and (d) $B(E2)$. The double pointed arrows between the experimental
   doublet bands in (a) indicate their energy differences.}
   \label{observable}
\end{figure}

For config1, there appears two degenerate minima ($|\varphi_{\min}|
\sim 15^\circ$) on the Routhian curve at $\hbar\omega=0.50$ MeV. However,
the curve is rather flat in the region $-20^\circ \leq \varphi \leq 20^\circ$.
For higher $\hbar\omega$, the barrier of the Routhian at $\varphi=0$ increases
rapidly. The height of the barrier goes up to $\sim 150$ keV at $\hbar\omega=0.60$ MeV.
The higher and wider barrier of the Routhian will suppress more the
tunnelings between the left- and right- handed configurations and lead
to a stronger degeneracy of the chiral doublet bands~\cite{Q.B.Chen2013PRC}.
This is in qualitative agreement with the experimental observation
of the closer of the bands D3 and D4 shown later in Fig.~\ref{observable}(a).

For config2, unfortunately, the calculations could not be followed successfully
up to the largest observed spin $23.5\hbar$. Convergent results of configuration-fixed
calculations could be attained only up to $\hbar\omega=0.55$ MeV, corresponding
to $I\sim 21\hbar$. By further increasing $\omega$, a level crossing between
the neutron $h_{11/2}$ and $h_{9/2}$ orbits appears. This leads to a new
configuration $\pi h^2_{11/2}\otimes \nu h^{1}_{9/2}$. Therefore, it might
imply that the $\omega_{\textrm{crit}}$ for config2 in the present 3D-TAC CDFT
calculation is larger than $\hbar\omega=0.55$ MeV. For comparison, the
$\omega_{\textrm{crit}}$ is $\hbar\omega=0.50$ MeV in the 3D TAC calculation
based on hybrid of Woods-Saxon and Nilsson potential combined with
the shell correction method in Ref.~\cite{S.Zhu2003PRL}.

\begin{figure}[!ht]
\includegraphics[width=8.0 cm]{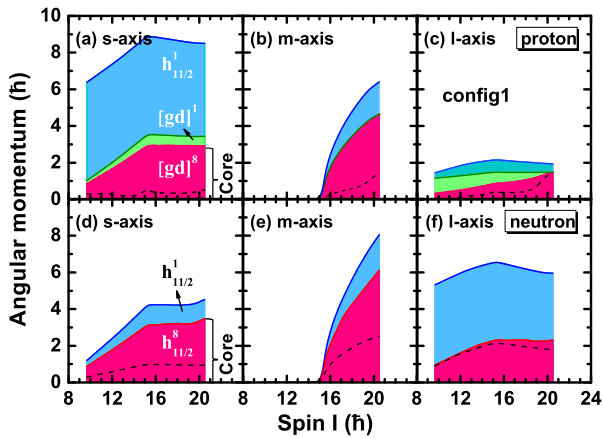}
  \caption{Contributions of the valence protons and
   neutrons in the $h_{11/2}$ and $(gd)$ shells as well as the residual core-like part
   to the total proton and neutron angular momenta along the $s$-,
   $m$-, and $l$- axis for config1 in $^{135}$Nd.}
   \label{jcom}
\end{figure}

The rotational motion of triaxial nuclei attains a chiral character
if the angular momentum has substantial projections on all three
principal axes of the triaxially deformed
nucleus~\cite{Frauendorf1997NPA}. In a microscopic picture, the
angular momentum comes from the individual nucleons in a
self-consistent calculation. In order to check the mechanism behind
the generation of the angular momentum, it is important to extract
the contributions of the individual nucleons to the angular
momentum. The results from 3D-TAC CDFT for config1 at the minima of
the Routhian are presented as an example in Fig.~\ref{jcom}.

One observes that the contribution to the proton angular momenta along
the $s$-axis originates mainly from the high-$j$ orbit, i.e., a proton
filling at the bottom of $h_{11/2}$ orbit. The valence proton occupying
in the $(gd)$ shell orbit contributions quite small along the three
principal axes ($<1\hbar$). In contrast, a neutron sitting at the
third $h_{11/2}$ shell from the top downward contributes
an angular momentum of roughly $4.5\hbar$ along the $l$-axis.
When the rotational frequency increases, the contributions of the
valence protons and neutron change barely along the $s$- and $l$- axis,
respectively.

The equivalent ¡°core¡± composed from the remaining nucleons now can be
separated as two parts, i.e., proton and neutron parts. In more detail, the eight
protons/neutrons in the $(gd)$/$h_{11/2}$ shell contribute to the
$s$-axis component, and the other nucleons make main contributions
on the $l$-axis component. Both proton and neutron parts contribute
together to the $m$-axis component after the $\omega_{\textrm{crit}}$.

\begin{figure}[!ht]
\includegraphics[width=8.0 cm]{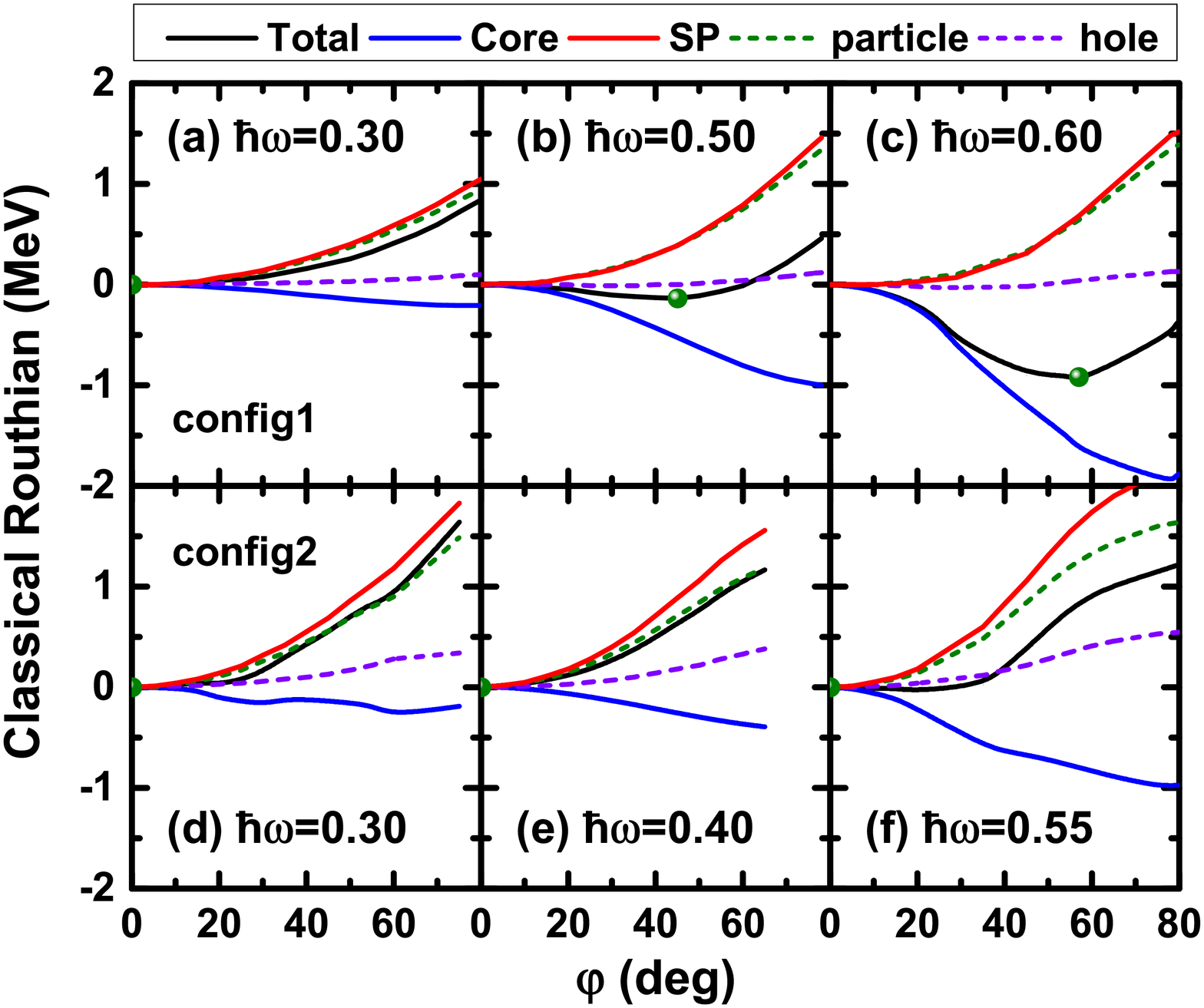}
   \caption{Classical Routhians as functions of $\varphi$ for the total, core
   as well as the valence particle(s), valence hole(s), and their summation (SP).
   All curves are normalized at $\varphi=0$. The minimum
   of the total Routhian is denoted by the ball.}
   \label{CR}
\end{figure}

Adopted the same ideas in Refs.~\cite{J.Meng1993APS,
Frauendorf1996ZPA, Frauendorf1997NPA}, the model study separates the
nucleus into several valence nucleons and a collective core-like
part. Correspondingly, the total classical Routhian of the nucleus
is calculated as
\begin{align}\label{eq1}
  E_{\rm{Total}}^\prime
   &=E_{\textrm{SP}}^\prime+E_{\textrm{core}}^\prime
    =\sum_{i=\textrm{p,h}}\epsilon_i^\prime
   -\frac{1}{2}\sum_{k=s,m,l} \mathcal{J}_k\omega^2_k,
\end{align}
in which the first term $\epsilon_{\textrm{p,h}}^\prime$ is the single-particle
Routhian for the valence particle(s) and hole(s), and the second term the
Routhian for the core-like part. Inspired by this expression, we make an attempt
on understanding the appearance of tilted rotation shown in Fig.~\ref{Routhians}.
Starting from 3D-TAC CDFT results, we extract, intuitively, the values of
$\epsilon_{\textrm{p,h}}^\prime$ in Eq.~(\ref{eq1}) from the obtained
single-particle Routhian for those valence particle(s) and hole(s) that
the configuration information gives. As a consequence, the ``core'' is
composed by the remaining nucleons. Its angular momentum $R_k^{\textrm{core}}$
is thus the summation of these remaining nucleons' angular momentum.
Correspondingly, its MoIs can be extracted as
$\mathcal{J}_k=R_k^{\textrm{core}}/\omega_k$, and finally its
Routhian $E_{\textrm{core}}^\prime$ is yielded by Eq.~(\ref{eq1}).
The obtained results of these classical Routhians as functions
of $\varphi$ are shown in Fig.~\ref{CR}.

It is observed that the values of $\epsilon_{\textrm{p,h}}^\prime$ and
$E_{\textrm{core}}^\prime$ increase and decrease with $\varphi$,
respectively. The former indicates the valence particle(s) and
hole(s) prefer to the $sl$-plane, and the latter represents that
the core-like part favors the $m$-axis. Their competitions determine the
rotational orientation of the total system. These features are
consistent with previous model studies~\cite{Frauendorf1997NPA}.

For config1, after $\hbar\omega=0.50$ MeV, the steeper variation behavior of
$E_{\textrm{core}}^\prime$ than those of $E_{\textrm{SP}}^\prime$ provides
stronger Coriolis force and drives rotational axis deviating from the $s$-$l$ plane
to minimize the energy. The minima of total classical Routhian thus shift from zero
to nonzero, which corresponds to the aplanar solution, in line with the results
shown in Fig.~\ref{Routhians} though the detailed value of $\varphi_{\min}$
and the height of the barrier are different. Therefore, a transition
from planar to aplanar rotation has been displayed for config1.

For config2, the increments in $E_{\textrm{SP}}^\prime$ are larger than the
decrement in $E_{\textrm{core}}^\prime$. This is mainly caused by the two
aligned $h_{11/2}$ particles, which has a very large alignment along the
$s$-axis and make their Routhians become very steep with the increase of
$\varphi$. The minimum of total classical Routhian stays at $\varphi_{\min}=0$,
which corresponds to the planar regime. This is also agreement with the microscopic
results shown in Fig.~\ref{Routhians}.

In the previous model studies~\cite{Frauendorf1997NPA, B.Qi2009PLB, Q.B.Chen2018PLB,
Q.B.Chen2019PRC, B.F.Lv2019PRC}, the transition from planar to aplanar rotation
is attributed to the MoI of the $m$-axis, with the assumption
of irrotational flow type $\mathcal{J}_k^{\textrm{irr}} \propto \sin^2(\gamma-2k\pi/3)$,
is the largest. In this work, the three MoIs for the core-like part are
extracted as $\mathcal{J}_k =R_k^{\textrm{core}}/\omega_k$ ($k=s,m,l$) from the
3D-TAC CDFT results. We find that $\mathcal{J}_k$ does not change much with the
increase of $\omega$ and $\varphi$. In addition, $\mathcal{J}_m$ is indeed
the largest for both configurations. In detail, e.g., at $\hbar\omega=0.30$ MeV,
the ratio of $\mathcal{J}_m:\mathcal{J}_s:\mathcal{J}_l$ at $\varphi=0$ is
about $1.00:0.65:0.50$ for config1 and $1.00:0.78:0.76$ for config2,
respectively. This feature is consistent with those empirical MoI values
extracted from the $2_2^+$ states in even-even nuclei~\cite{Allmond2017PLB}
and supports the assumption of $\mathcal{J}_k^{\textrm{irr}}$ in the model study.
In Ref.~\cite{Olbratowski2004PRL} the MoIs are extracted for the whole system
from the 3D-TAC Skyrme-Hartree-Fock calculations, and the $m$-axis one
is the largest as well. Furthermore, one notes that the ratios of
$\mathcal{J}_s(\mathcal{J}_l) :\mathcal{J}_m$ for config2 is larger
than that for config1. This makes, according to the formula of
$\omega_{\textrm{crit}}$ in Ref.~\cite{Olbratowski2004PRL}, the planar
rotation become more stable, and the aplanar rotation is not easy
to be obtained as shown in Fig.~\ref{Routhians}.

Therefore, we can possibly understand how dose the titled rotation occur
in the 3D-TAC calculations. Though moving in the same rotating mean field, some
(unpaired) valence high-$j$ nucleons are active with significant single-particle
motion and act as valence particle(s) or hole(s), while the (paired) others are not
that active but exhibit collective behavior and form a stable core-like part. This core-like part
has finite MoIs along the three principal axis (as indication of triaxial deformation),
and the $m$-axis one is the largest. At low $\omega$, the collective core angular
momentum is small and is driven to lie in the $sl$-plane by the strong Coriolis forces
from the valence particle(s) and hole(s). With increasing $\omega$, the
gradual increasing core angular momentum along the $m$-axis becomes comparable
to those of valence particle(s) and hole(s) along $s$- and $l$- axis and
results in the transition from planar to aplanar rotation.


With the above studies, the comparisons between the 3D-TAC CDFT results
and the available experimental data~\cite{S.Zhu2003PRL, Mukhopadhyay2007PRL,
B.F.Lv2019PRC} are given in Fig.~\ref{observable}.

As the calculations are carried out without additional adjustable parameters
and the theoretical bandhead energies of both configurations are shifted
by the same value, the qualities of present reproduction are reasonable for
energy spectra. In addition, the larger experimental energy differences
between bands D6 and D5 than bands D4 and D3 at low spin region correlate
with their steeper behavior of Routhian as observed in Figs.~\ref{Routhians}
and \ref{CR}.

The calculated $B(M1)/B(E2)$ values of config1 show a steep falling behavior
in the planar rotation regime, and the falling tendency slows down after
the transition to aplanar rotation. There is only one experimental value of
$B(M1)/B(E2)$ at $I=14.5\hbar$ for band D3~\cite{B.F.Lv2019PRC}, which is
slightly smaller than those of config1. For band D5, one observes that the
calculated result with config2 shows a good agreement with the data within the
error bar after $I=18.5\hbar$. However, it overestimates the data in the lower
spin region. It is caused by, as shown later, underestimating the $B(E2)$
values from $I=14.5$ to $18.5\hbar$.

For $B(M1)$, results of config2 are in reasonable agreement with the available
data of band D5. They show a smooth decreasing tendency with spin. The config1,
as less of a high-$j$ $h_{11/2}$ particle than config2, gives smaller $B(M1)$
values. However, it also shows a decreasing trend in the planar rotation region.
In the aplanar region, it increases a bit.

The $B(E2)$ value depends on the deformation parameters and the
orientation angles $\theta_{\min}$ and $\varphi_{\min}$. As the
deformation and $\theta_{\min}$ change slightly in config2, the
calculated $B(E2)$ values are roughly constant, in line with the
behavior of the experimental value. However, the calculated values
of $B(E2)$ are somewhat smaller than the observed values. As shown
in Ref.~\cite{P.W.Zhao2015PRC}, the including of pairing
correlations could improve a bit this agreement. For config1, the
calculated $B(E2)$ values increase slowly for planar rotation while
rapidly for aplanar rotation. This implies that, as shown in
Figs.~\ref{jcom} and~\ref{CR}, the collective core-like part plays
more and more important role with the increase of spin. Thus further
experimental efforts in particular the lifetime measurement to
extract $B(M1)$ and $B(E2)$ values, are encouraged for the bands D3
and D4.


In summary, the 3D-TAC CDFT is used to study the chiral modes in
$^{135}$Nd. The transition from planar to aplanar rotation is found
for config 1 of band D3, while only the planar rotation for config 2
of band D5. By modeling the motion of the nucleus in rotating mean
field as the interplay between the single-particle motion of several
valence particle(s) and hole(s) and the collective motion of a
core-like part, a classical Routhian is extracted. This classical
Routhian gives qualitative agreement with the 3D-TAC CDFT result of
the critical frequency for the transition from to aplanar rotation.
In addition, the extracted MoIs of the core-like part has the
largest value along the $m$-axis. Based on this investigation a
possible understanding of tilted rotation appearing in a microscopic
theory is provided. Current study also indicates the effects of
core-like part indeed exists in the realistic rotating nucleus and
thus provides a justification for the assumptions of the
phenomenological model, and more importantly reinforces the duality
of single-particle and collective properties in atomic nuclei.

\begin{acknowledgments}

The authors thank P. W. Zhao for providing the numerical code of
3D-TAC CDFT. This work is supported by the National Natural Science
Foundation of China (NSFC) under Grants No. 11775026 and 11875027,
and the Deutsche Forschungsgemeinschaft (DFG) and NSFC
through funds provided to the Sino-German CRC110 ``Symmetries and
the Emergence of Structure in QCD'' (DFG Grant No. TRR110 and NSFC
Grant No. 11621131001).

\end{acknowledgments}


\end{document}